# Ultra-Wideband Communications: Interference Challenges and Solutions

Brian Nelson, *Member, IEEE*, Hussein Moradi, *Member, IEEE,*
and Behrouz Farhang-Boroujeny, *Life Senior Member, IEEE*

*Abstract*—The idea of ultra-wideband (UWB) communications for short ranges (up to a few tens of meters) has been around for nearly three decades. However, despite significant efforts by the industry, UWB deployment has not yet reached its predicted potential. This article, thus, seeks to rectify this situation by providing a practical examination of UWB interference conditions. Through a spectrum survey of today's wireless environments, we explore the interference that UWB devices may face from a perspective of outage probability in both high- and low-rate configurations. We find that by suppressing interference, the outage probability can be reduced by one or more orders of magnitude. In the non-line-of-sight channels, in particular, we find that both interference suppression and bandwidth expansion are required to support the minimum data rates suggested in the IEEE802.15.4 series of standards. We connect these findings to a recently proposed UWB signaling method based on filter banks and show this method fulfills the above requirements for implementing effective UWB systems.

## I. INTRODUCTION

Ultra-wideband (UWB) systems provide unique opportunities for short range integrated sensing and communication (ISAC). Commercial access to this technology is growing—in 2022 approximately 21% of all new smartphone shipments were UWB capable, with significant growth anticipated over the next 5 to 10 years [1]. UWB signals are characterized by a very large transmission bandwidth, $\geq$ 500 MHz, or $\geq$ 20% of the signals center frequency. UWB transmissions have been approved for unlicensed use by a number of spectrum-regulating bodies, including the Federal Communications Commission (FCC), in the United States (US), as long as the transmitted power spectral density (PSD) satisfies a spectral mask provided by the regulating body of the region of operation. In particular, in the US, this spectral mask allows unlicensed transmission from $3.1 - 10.6$ GHz, as long as the PSD of transmitted power out of the antenna remains below $-41.3$ dBm/MHz [2].

Beside UWB, many methods for facilitating ISAC devices have been presented in the literature, including multi-modal sensing. Because ranging and sensing accuracy depends on the occupied bandwidth, any ISAC design will require a large bandwidth. One proposed approach for accessing sufficient bandwidth is through the use of mmWave designs [3], [4]. Though these bands facilitate precise sensing, several challenges exist. These include a high propagation loss, particularly through obstructions. To compensate for this loss, massive multiple input multiple output (mMIMO) arrays are suggested [3], but these result in increased computational complexity and hardware costs [5]. Additionally, in the mmWave bands, the channel experiences a high degree of channel nonstationarity [4], resulting in more overhead and complexity for channel estimation and tracking.

UWB provides an interesting alternative to mmWave ISAC—UWB systems occupy a large enough bandwidth to facilitate precise localization and sensing, but at a low enough center frequency that signal propagation loss is significantly less. Additionally, because the center frequency is lower, channel Doppler (viz., variation in time) is less severe and, hence, less overhead for channel estimation and tracking is required. However, the spectral overlap of UWB signals with the licensed and unlicensed users creates serious challenges. These challenges are exacerbated by the fact that the UWB signal at any receiver input often drops significantly below the

This manuscript has been authored by Battelle Energy Alliance, LLC under Contract No. DE-AC07-05ID14517 with the U.S. Department of Energy. The United States Government retains and the publisher, by accepting the article for publication, acknowledges that the United States Government retains a nonexclusive, paid-up, irrevocable, world-wide license to publish or reproduce the published form of this manuscript, or allow others to do so, for United States Government purposes. STI Number: INL/CON-24-81301.

Brian Nelson and Hussein Moradi are with the Idaho National Laboratory, Idaho Falls, ID 83415, USA.
Behrouz Farhang-Boroujeny is with Electrical and Computer Engineering, University of Utah, Salt Lake City, UT 84112, USA.

equipment noise and interfering signals may be a few orders of magnitude above the equipment noise. It turns out that without an effective method of interference suppression, many potential benefits of UWB may not be realizable. Given the small size of the interfering signal bandwidths relative to that of UWB signals, in the rest of this article, we refer to the former as narrow-band interference (NBI).

In the early days of UWB technology, the focus was primarily on achieving very high bit rates over short distances, driven by its potential for high-speed wireless personal area networks (WPANs). It was argued that the extremely wide frequency bands should allow data rates of at least a few hundred mega-bits per second. Early research and standardization efforts, such as the IEEE 802.15.3a task group, aimed to use UWB for applications like wireless USB, high-definition video streaming, and cable replacement. Despite the promise of high bit rate and precise ranging capabilities, UWB faced challenges such as regulatory constraints, coexistence with narrowband systems, and market competition from technologies like Wi-Fi and Bluetooth, which limited its early adoption in consumer electronics.

The subsequent follow up by the industry led to IEEE802.15.4 series of standards. These standards are focused primarily on ranging with the inclusion of data rates significantly lower than the original rates that were envisioned in IEEE802.15.3a (up to 500 Mb/s). Despite these set-backs, hope for development and standardization of higher rate UWB systems remain. High-rate designs would facilitate the realization of a more capable UWB ISAC design. Reference [6] presents an excellent history of the first few years after the FCC allocation of the 7.5 GHz of spectrum for UWB communications and what happened at the end. In pursuit of these objectives, we explore the potential for a high-rate UWB communication system design.

To develop an effective high-rate and ISAC UWB system, consideration of the interference environment is essential. Previous works, such as [7], have relied on general spectrum surveys such as [8] to characterize the interference environment. Other works have demonstrated NBI suppression algorithms in contrived interference scenarios [9], or discussed interference suppression techniques without directly considering the likely interference environment [10]. Additionally, these studies were performed in the early 2000's, when mobile wireless technologies were not as prevalent as they are today. For these reasons, we believe that a fresh study of current interference environments and their impacts can shed valuable insights into the design and simulation of the next generation UWB transceiver systems.

In this article, we first present the results of an extensive spectrum survey to characterize the wireless environment and the level of interference that may be seen in some common UWB environments over the low bands of the IEEE802.15.4 standard. We explore the impacts of NBI from a water-filling capacity perspective. The water-filling formulation allows for a direct comparison of the theoretical capabilities of a UWB system equipped with an optimal NBI suppression scheme and a naive receiver that views NBI as part of the channel noise. This study confirms that, in real-world environments, an effective suppression of NBI can reduce the outage probability by many orders of magnitude and demonstrates that, to achieve high data rates, advanced signal processing methods for removing the NBI is necessary. Next, following the proposal in [11], we demonstrate that a filter bank multi-carrier spread spectrum (FBMC-SS) design naturally fits the water-filling capacity formulation. We show the effectiveness of this design by using the captured noise and interference to impair the transmission. It is seen that the filter bank design is able effectively suppress the interference and approach the theoretical outage probabilities to within an expected implementation loss.

The rest of this paper is organized as follows. In Section II, we present our experimental setup and the radio frequency (RF) environments sensed. In Section III, we present the parameters used for our spectrogram computation and the analysis methods used for computing the outage probabilities. The theoretical outage probability results for two configurations of the receiver, one with interference suppression and one without interference suppression, are presented in Section IV. In Section V, we discuss the use of filter bank multicarrier method as an effective method for NBI suppression and an implementation that can realize the full potential of UWB systems. Discussion and concluding remarks are presented in Section VI.

## II. SPECTRUM SURVEY

### A. Hardware Setup

We use a set of Ettus X310 universal software radio peripherals (USRPs). For our spectrum sensing purpose, each USRP contains two independent RF channels that can be tuned to different carrier frequencies. Each channel is sampled at 200 MHz, with a 160 MHz anti-aliasing filter passband bandwidth. These USRPs are chosen because they allow for tight timing synchronization across different RF channels within and across different USRPs by using a synchronization module. By tuning the channels to separate carrier frequencies and synchronizing the RF channels in time, a wide bandwidth can be sensed simultaneously. This implementation follows the idea brought up in [11], where to avoid the use of highly inefficient very high-speed digital to analog converter (DAC) and analog to digital converter (ADC), the UWB band is divided to a set of subbands, each having their own DAC and ADC, as well as radio chains.

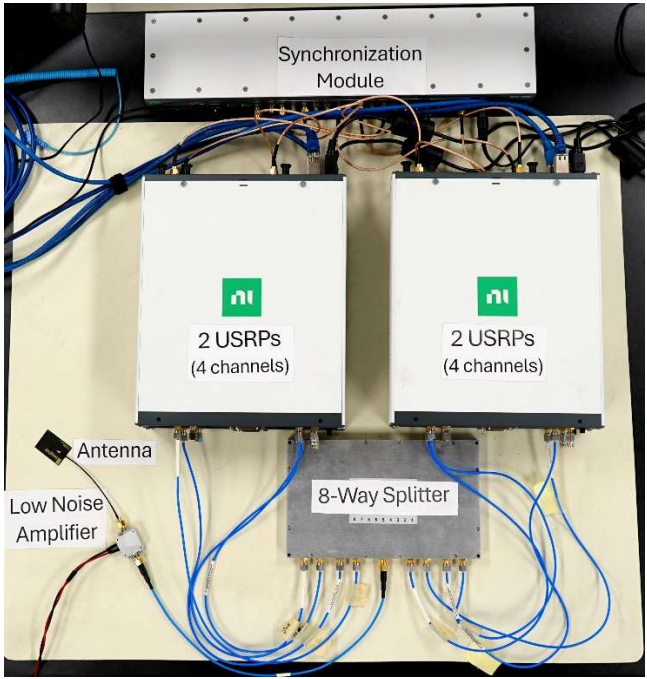

Fig. 2: The hardware sensing setup. Each Ettus X310 USRP provides two radio chains covering 2×160 MHz of spectrum. All radio chains on the four USRPs are time and frequency synchronized through the synchronization module.

In our setup, we use 4 concatenated USRPs. The USRP radio channel center frequencies are separated by 160 MHz, leading to a sensing bandwidth (8 × 160 MHz = 1.28 GHz), in total. Moreover, since X310 USRPs are only tunable to carrier frequencies below 6 GHz, our measurements are limited to the low band channels of IEEE802.15.4. Fig. 1 depicts our hardware sensing setup.

### B. Received Signal Power

To place the sensed interference power in the context of a UWB signal, we have calibrated the radios so that the received signal power will be known in dBm per unit of bandwidth. We compare this power to the received signal power following the IEEE802.15.4 channel path loss model [12]. By studying the signal to interference plus noise ratio (SINR) at different parts of the UWB spectrum, we evaluate the outage probabilities of UWB for the case of a simple detector and for the case of a receiver that is equipped with an effective interference suppressor.

The IEEE802.15.4 channel model, [12], provides a general UWB path-loss (PL) model, as well as parameters for a number of operating environments, such as, residential, office, outdoor, and industrial channels. These models also provide parameters for line-of-sight (LOS) and non-line-of-sight (NLOS) channels. This path loss model was created based on an extensive set of measurement surveys. See [12] and the references therein for detailed descriptions of the PL for each environment. Using this model, the received signal power at frequency $f$ and distance $d$ is obtained as

$$P_{rx}(f) = -41.3 + 10 log_{10} PL(d,f) \text{ dBm/MHz}, \quad (1)$$

where the first term on the right-hand side comes from the spectral mask.

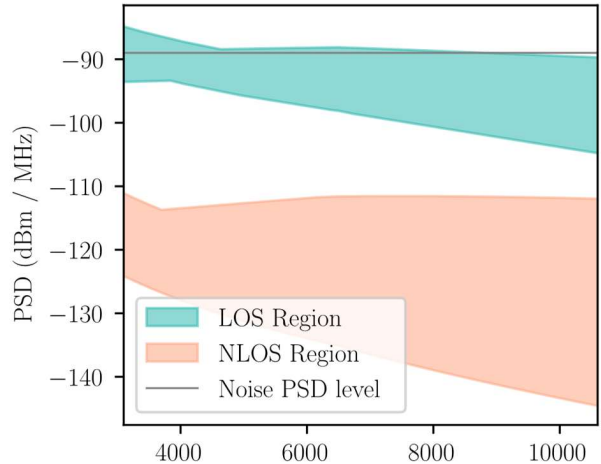

Fig. 1: PSD of the equipment noise and those of the UWB received signal for both LOS and NLOS channels. PSD ranges shown covers the distances of 3 to 8 meters.

Fig. 2 presents a diagram showing the PSD of our equipment noise (as what one may encounter in a typical radio), along with the ranges of the received signal PSD for a set of LOS and NLOS channels based on the IEEE802.15.4a channel model. These are the combined results in an office setting and an industrial site for transmission ranges of 3-8 meters. The equipment noise is obtained from the USRP Ettus data sheets and also confirmed through our measurements. The important point to note here is that the received UWB signal is mostly at some negative signal-to-noise-ratio (SNR) values. For NLOS channels, in particular, the SNR range is 25 dB or more below the equipment noise. It should also be noted that the equipment noise presented in Fig. 2 is averaged over the spectral range. Typical radio transceivers have a noise figure that is dependent on tuned center frequency. This is due to differences in the performance of analog components in the receiver processing chain at different center frequencies. As a result, the outage probabilities presented later are dependent on which part of the spectrum is being utilized for transmission.

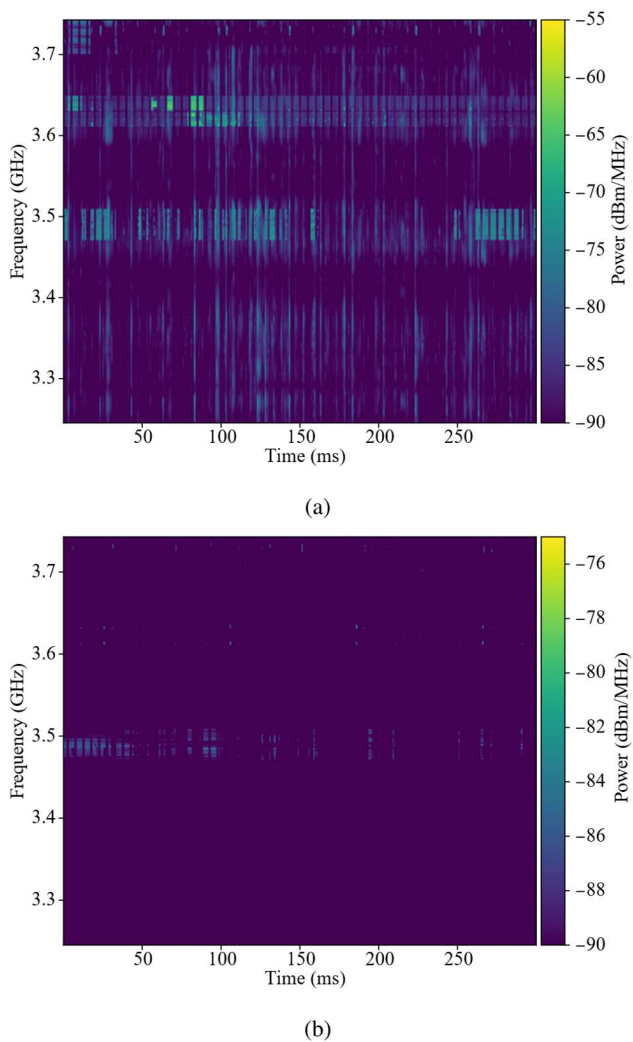

Fig. 3: Examples of spectrograms captured signals in IEEE802.15.4 channel 1. (a) A busy scenario. (b) A quiet scenario.

### C. Measurement Environments

The environments for our measurements are chosen to address a variety of use cases of UWB system. The surveys were performed at a number of locations in Salt Lake City, Utah. We study the interference scenarios both in indoor and outdoor environments. The locations and environments were selected to provide a range of spectrum utilization scenarios typical consumer devices are likely to experience.

As may be expected, some operating environments experience only a small amount of spectral activity, while other environments experience frequent, significant partial band interference. Because a user is likely to consistently operate a device in the same environment, e.g. at their home or office space, they will often experience the same conditions. If a user's home or office happens to be located near a cellular base station, or in a location with significant Wi-Fi activities (say, in an apartment complex), they will always experience a harsh interference environment. Due to the ubiquitous and growing demand for Wi-Fi and cellular mobile access, these locations are common and are only likely to increase in next generation systems. For this reason, we have focused our attention on the environments with significant interference to provide insights into the typical experience of these users.

We collected spectral activities at the following locations:

- Inside the University of Utah Marriott library. The general observation here was a lot of interference from Wi-Fi systems spread around the building. Very little cellular signal could be observed, given that the building has thick concrete walls, hence, shielded from the outdoor spectral activities.
- In the University of Utah's Merrill Engineering Building. The signal activities in this building were somewhat similar to those in the Marriott library, with somewhat smaller Wi-Fi activities.
- In an outdoor, urban environment. Here we started collecting data at a point close to a base station tower and walked away to a distance of about half a mile from the tower. We believe these data are representative of what may be observed in residential homes at different distances from the cellular towers. This assumption is reasonable because residential homes are often made of materials that are transparent to cellular signal activities. Therefore, the data from these surveys will be representative of the interference environment UWB devices will experience in these buildings. Of course, within residential homes, Wi-Fi signal activities may further pollute the spectrum.

Our spectrum survey follows Channels 1 through 4 of the IEEE802.15.4 standard. The survey channels have the following center frequency (CF) and bandwidth (BW):

- Channel 1: CF = 3.4944 GHz, BW = 499.2 MHz
- Channel 2: CF = 3.9936 GHz, BW = 499.2 MHz
- Channel 3: CF = 4.9928 GHz, BW = 499.2 MHz
- Channel 4: CF = 3.9936 GHz, BW = 1280 MHz

### III. ANALYSIS

#### A. Spectrograms

To analyze the captured data, we study the spectrogram of the received data. The spectrogram is calculated for each radio channel independently. The frequency bins corresponding to the transition band of the anti-aliasing filters are discarded, and the remaining frequency bins from different radio chains are concatenated across frequency, allowing for a view of the full UWB band.

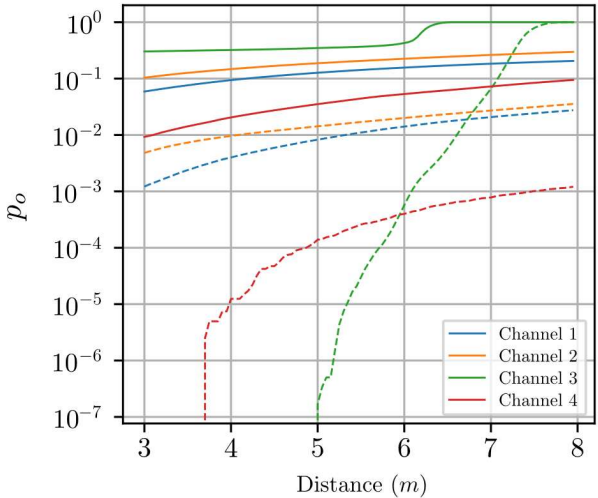

Fig. 4: Outage probability results over Channels 1 through 4 of IEEE802.15.4 standard in a LOS setup plotted against operating distance. Dotted and solid curves represent outage probabilities with and without interference suppression, respectively.

The spectrogram for each channel is computed by taking segments of 8192 samples of the collected data and applying a Blackman-Harris window, before applying a discrete Fourier transform. This results in a frequency resolution of 24.41 kHz and a time step of $\Delta t = 40.96$ μs. Examples of spectrograms from a couple of outdoor captures are presented in Figs. 3(a) and (b). Fig. 3(a) presents a scenario with a high-level of cellular activities in Channel 1 of the IEEE802.15.4 standard. Fig. 3(b), on the other hand, presents a scenario at a location with a lower level of primary user activities.

### B. Outage Probability

To determine the impacts of NBI on a hypothetical UWB transmission, we study the outage probability, defined as

$$p_{out} = \Pr(C < R), \quad (2)$$

where $C$ is the Shannon channel capacity and $R$ is the transmission rate.

To determine the impact of partial band interference on the outage probability, we compare two methods of computing the capacity. One method represents a system that performs no sophisticated interference suppression. It simply looks at the averaged SINR across the band of transmission $W$ and inserts the result in the Shannon capacity formula

$$C = W\log_2(1 + SNR). \quad (3)$$

This may be thought of as an approximation to some of the current implementations of UWB systems that rely on the collected signal energy in each epoch of symbol duration.

The second method that we consider in calculating the capacity segments the transmission band into a set of narrow bands over which the channel may be well approximated as an additive white Gaussian noise (AWGN) channel, hence, an accurate estimate of the capacity at each segment can be obtained by using the Shannon capacity formula. The total channel capacity is then obtained by summing the capacities from the individual segments. This scheme closely approximates the way the filterbank-based designs of [11] and [13] handle the NBI and parallels the conventional "water-filling" capacity results. For the sake of our reference in the rest of this paper, we refer to this method as '*a detector with interference suppression*' and refer to the method discussed in the previous paragraph as '*a detector without interference suppression*'. It can also be noted that with an ideal receiver and without NBI, the capacity of these schemes is the same. In our case, the capacities are approximately the same, because as with any physical radio hardware, the equipment noise PSD of the radio is not perfectly flat, which the interference suppression scheme takes advantage of to arrive at a slightly increased capacity. Though this difference is present, it has little impact on our overall findings.

Comparing these outage probabilities provides valuable insights into the impacts of NBI on a UWB system. It provides an upper bound on the performance of any system and provides insights into practical methods for approaching this bound. A transceiver design that partitions the spectrum into well-isolated bands and codes across them, such as the FBMC-SS design of [13], will be an effective way of minimizing the outage probability for a given transmission rate.

### IV. OUTAGE PROBABILITY RESULTS

In this section, we present the outage probabilities of the two methods discussed above for the data gathered through the spectrum survey. Here, we use the office environment PL model parameters when determining the outage probabilities. We present the result of a detector with interference suppression and a detector without interference suppression.

Fig. 4 presents theoretical outage probabilities for Channels 1 through 4 of the IEEE802.15.4 standard with and without interference suppression. Here, the PL is based on a LOS office environment, the bit rate is 124.75 Mbps, and the interval over which the outage condition was calculated was 40.96 μs (approximately the duration of a 5120 bit packet). While for Channels 1 and 2, the interference suppression results in one to two orders of magnitude improvement, the improvements for Channels 3 and 4 are much larger. For Channel 3, in particular, while the performance without interference suppression is very poor, interference suppression leads to a significant drop in the outage probability for distances of less than 6 meters.

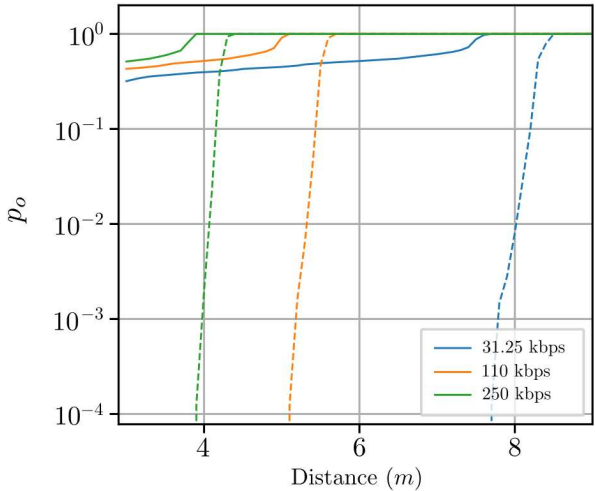

Fig. 5: Outage probability results for several low rates of the IEEE802.15.4 standard over Channel 3 of IEEE802.15.4 standard in a NLOS setup plotted against operating distance. Dotted and solid curves represent outage probabilities with and without interference suppression, respectively

The results of Channel 4 in Fig. 4 also deserve some special attention. Recall that while Channels 1, 2, and 3 have the minimum bandwidth of 500 MHz, the Channel 4 bandwidth is wider, but with the same center frequency as Channel 2. The larger bandwidth results in a greater processing gain and, hence, leads to an improved outage probability for schemes with and without interference suppression due to the extra processing gain and degrees of freedom offered by the wider bandwidth of Channel 4. Without interference suppression, the wider bandwidth results in only a small improvement over the narrower band channels. In fact, when interference suppression is applied to the narrower band channels, the performance exceeds the wideband, no suppression design. This demonstrates that bandwidth expansion alone should not be pursued as a method to improve performance across all user scenarios. It implies that in a practical system, if the design complexity can only afford to either suppress interference or expand bandwidth, NBI suppression will lead to more reliable performance. When a wider bandwidth transmission is used, more un-interfered portions of the spectrum are available. In contrast to the scheme without suppression, the interference suppressor uses these to support more reliable communication. In addition, we see that for high-rate communications, 500 MHz of bandwidth is likely insufficient for reliable performance even with interference suppression. We see both NBI suppression and bandwidth expansion are required for good performance. For this reason, flexible suppression schemes that scale to wider transmission bandwidths are required. The FBMC-SS designs presented in [11], [13] provide such a solution. In Section V, we present some results to further demonstrate the effectiveness of such designs.

Fig. 5 presents a set of outage probability results over a NLOS channel. Given the very low SNR of NLOS channels, the data rates that may be supported in this scenario are relatively low. Hence, the data rates that are considered for our study, here, are the three lowest rates in the IEEE802.15.4 standard; namely, the data rates of 31.25 kbps, 110 kbps, and 250 kbps [14]. The propagation loss was found using the NLOS office parameters, and interference was from all indoor captures over Channel 3 of IEEE802.15.4 standard. Recall that Channel 3 has a bandwidth of 500 MHz. The duration of each packet was chosen to be 30 ms (approximately, the duration of a 2560 bit packet at 110 kbps, with 25% preamble sequence overhead).

As seen in Fig. 5, a detector without interference suppression performs very poorly. At all rates and distances, the outage probability exceeds 40. This demonstrates that reducing the transmission rate without interference suppression does not provide a meaningful improvement to the robustness of the link. On the other hand, a receiver equipped with an interference suppressor, performs significantly better if the SNR of the uninterfered portions of the spectrum is high enough to support the transmission bit rate. At longer distances, even with interference suppression, the SNR drops below the channel capacity limit, hence, the outage probability approaches 1. These results demonstrate that for reliable low-rate communications, interference suppression is required. To increase the range for a fixed bit rate, more bandwidth can be used, similar to what was observed for the high-rate design. For the FBMC-SS design proposed in [11], this can be accomplished by using a larger number of moderate rate radio chains tuned to adjacent center frequencies.

## V. FILTER BANK MULTICARRIER

To approach the theoretical outage probabilities predicted by the captured spectrum surveys, we propose using the FBMC-SS design suggested in [11]. In this design, a filter bank with a well-designed prototype filter is used to partition the spectrum into a set of well-isolated subcarrier bands. Information is coded across these subcarrier bands, and an optimal combiner is used to suppress interference and recover the transmitted information.

Based on the interference surveys, we see that using narrow subcarrier bands to precisely suppress NBI will lead to the highest performance. As observed in [13], reducing the bandwidth of the filter bank subcarrier bands

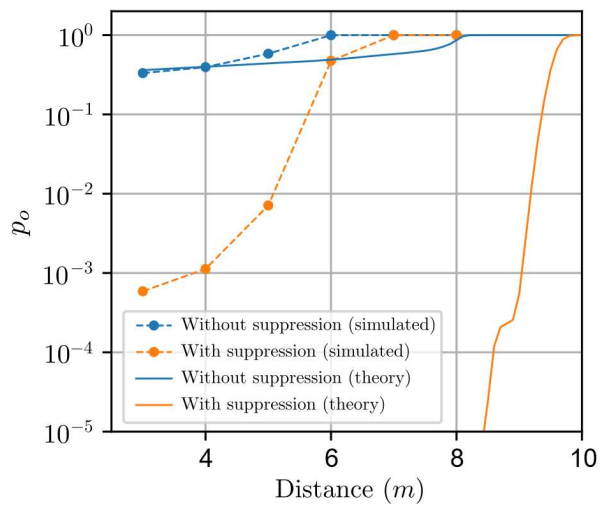

Fig. 6: Comparison of ideal theoretical outage probabilities and those obtained from a simulated system using a realistic channel code and an advanced filterbank-based signal processing method for interference suppression.

results in a reduced peak bit rate for an FBMC-SS design. In [13], a method for addressing this is presented. Here, the subcarrier bands are chosen to be narrow, allowing for precise interference suppression, but are partitioned into subsets, or "data streams" that are coded independently. The subcarrier bands in a data stream are maximally separated in frequency to provide diversity against wider bandwidth interference events

In this section, we present the results of a simulation using this parallel data stream approach for a practical channel code and an advanced filter bank-based signal processing method for interference suppression. We compare the performance of a design equipped with an interference suppressor, and one without a suppressor. Details of this implementation can be found in [13]. The channel encoder is a 1/3 rate, 5G low-density parity-check (LDPC) code with information block lengths of 2560 bits [15, pp. 529-531]. The coded bit rate is 83.2 Mbps. In the simulation, realizations of the IEEE802.15.4 multi-path channel model for the LOS office environment are used to simulate the impacts of the multi-path environment. After passing through the simulated channel, the signal is scaled to have the appropriate received power for the considered distance and then segments of the spectrum survey noise and interference are added. In an effort to keep the simulation complexity manageable, only Channel 3 of the Marriott library data set is used. Here, high-powered partial band interference is experienced approximately 30% of the time. Additionally, to isolate the interference resilience performance, we provide the receiver with perfect channel state information. In a practical system, additional loss can be expected due to synchronization error.

Fig. 6 compares the simulated frame error rate and the theoretical outage probabilities. The simulated system follows the respective theoretical curves to within an expected coding and implementation loss. Our detector is effectively a linear equalizer with some maximum ratio combining across different subcarrier bands that result in some processing gain; see [13] for details.

The FBMC-SS design demonstrates that by designing with the interference environment in mind and leveraging a waveform with a high degree of flexibility for suppressing interference, the system capacity can be approached. On the other hand, a design that does not consider the interference environment will have poor performance.

## V. CONCLUSION

This article proposed UWB as a candidate for ISAC systems in the next generation, short range wireless systems. Recognizing that NBI is a unique and significant challenge for UWB ISAC systems, and that wireless activities have changed significantly since prior surveys were completed in the early 2000's, we performed a number of surveys to quantify the level of interference that UWB devices may face in today's environments. The optimal interference suppressor was found to have the form of an FBMC-SS design, where the flexibility to suppress NBI could result in improved outage probability.

Using the outage probability formulation, we came to a few important realizations. From the perspective of a high-rate UWB design, an interference suppressor will result in a more substantial improvement in reliability than bandwidth expansion. It was observed that a bandwidth larger than 500 MHz and interference suppression will be required for reliable, high-rate performance. From the perspective of low-rate, NLOS designs, it was observed that reducing the transmission rate without NBI suppression had little improvement on the reliability.

From these results, we found that for the next generation, high-reliability UWB designs, new transceiver architectures will be required. To this end, we explored the performance of a practical UWB design based on FBMC-SS that allows for excellent, highly flexible interference suppression and straightforward scalability to wider bandwidths. It was shown that the performance of this design, when impaired by a simulated UWB multi-path channel and the interference from our spectrum surveys, followed the performance suggested by the outage probability results to within an acceptable coding and implementation loss.

ACKNOWLEDGEMENTS

This research made use of the resources of the High Performance Computing Center at Idaho National Laboratory, which is supported by the Office of Nuclear Energy of the U.S. Department of Energy and the Nuclear Science User Facilities under Contract No. DE-AC07-05ID14517.

BIOGRAPHIES

BRIAN NELSON received the B.S. Summa Cum Laude in computer engineering from Brigham Young University in Provo, UT, USA in 2021 and an M.S. degree in electrical engineering from the University of Utah in Salt Lake City, UT, USA in 2024. He is currently pursuing a PhD degree in electrical engineering from the University of Utah. His PhD research has focused on spread spectrum and ultra-wideband communications. He was an engineer in the modem group at L3Harris Technologies from 2021 to 2022. Since 2022, he has worked as a researcher in the Idaho National Laboratory Wireless Research Department in Salt Lake City, UT, USA.

HUSSEIN MORADI joined Idaho National Laboratory in November 2009. Dr. Moradi was declared "The Inventor of the Year" in 2017-2018 by the Idaho National Laboratory and Battelle Energy Alliance. Dr. Moradi earned his Ph.D. and Master's Degree from Southern Methodist University (SMU) Dallas, and his Bachelor Degree from University of Texas at Arlington (UTA); all of the degrees are in the EE discipline. He has been a member of the Board of Professional Engineers since the early 1990's. Prior to joining INL, Dr. Moradi assumed lead R&D responsibilities at NEC America, VeriFone and Kyocera.

BEHROUZ FARHANG-BOROUJENY received the Ph.D. degree from Imperial College, University of London, UK, in 1981. He is a professor at the University of Utah and prior to that he was with the National University of Singapore. He is an expert in the general area of signal processing for communications and has numerous contributions in the domains of adaptive filters, MIMO detection methods, and spread spectrum techniques. He is the author of the books "Adaptive Filters: theory and applications", John Wiley & Sons, 1998 and 2013 (second edition) and "Signal Processing Techniques for Software Radios", self-published at Lulu publishing house, 2009 and 2010 (second edition).